\documentclass[jkps,twoside,twocolumn,showpacs,showkeys]{revtex4}
\usepackage{graphicx}
\usepackage{amssymb}
\usepackage{amsmath}
\usepackage{bm}
\usepackage{multirow}
\usepackage[titletoc]{appendix}

\begin{document}
\setcounter{page}{0}
\title[]{Identification of Additional Jets in the $\ttbb$ Events by Using Deep Neural Network}
\author{Jieun \surname{Choi}}
\author{Tae Jeong \surname{Kim}}
\email{taekim@hanyang.ac.kr}
\author{Jongwon \surname{Lim}}
\author{Jiwon \surname{Park}}
\author{Yeonsu \surname{Ryou}}
\author{Juhee \surname{Song}}
\author{Soohyun \surname{Yun}}

\affiliation{Department of Physics, Hanyang University, Seoul}

\date[]{31 October 2019}

\newcommand{\mue}{\rm \mu^{\pm}e^{\mp}}
\newcommand{\emu}{\rm e^{\pm}\mu^{\mp}}
\newcommand{\ttcc}{\rm t\bar{t}c\bar{c}}
\newcommand{\ttbb}{\rm \MakeLowercase{t}\bar{\MakeLowercase{t}}\MakeLowercase{b}\bar{\MakeLowercase{b}}}
\newcommand{\ttbj}{\rm t\bar{t}bj}
\newcommand{\ttll}{\rm t\bar{t}LF}
\newcommand{\ttjj}{\rm t\bar{t}jj}
\newcommand{\ttH}{\rm \MakeLowercase{t}\bar{\MakeLowercase{t}}H}
\newcommand{\ttbar}{\rm t\bar{t}}
\newcommand{\bbbar}{\rm b\bar{b}}

\newcommand{\sttbb}{\sigma_{\rm t\bar{t}b\bar{b}}}
\newcommand{\sttjj}{\sigma_{\rm t\bar{t}jj}}

\newcommand{\GeVcc}{{\rm GeV}/c^2}
\newcommand{\GeVc}{{\rm GeV}/c}
\newcommand{\pt}{p_{\rm T}}
\newcommand{\mt}{m_{\rm T}}
\newcommand{\Ht}{p_{\rm T}}
    
\begin{abstract}
In the top quark pair production in association with the Higgs boson decaying to a b quark pair ($\ttH(b\bar{b}))$, the final state has an irreducible nonresonant background from the production of a top quark pair in association with a b quark pair ($\ttbb$).  
Therefore, understanding of the $\ttbb$ process precisely in particular differential cross-section
as functions of the properties of the additional b jets not from the top quark decay is essential
for improving the sensitivity of a search for the $\ttH(b\bar{b})$ process. 
The two additional b jets can be identified by using various approaches. In this paper, 
the performances are compared quantitatively in the lepton+jets decay channel in terms of 
the matching efficiency of assigning two additional b jets as a figure of merit.
We showed that a matching efficiency of around 40\% could be achieved using a deep neural network method. 
In the events with at least 4 b jets, this performance is 8\% better than that achieved using  minimum $\Delta R(b,\bar{b})$ method. This is consistent with the boosted decision tree method within its statistical uncertainty. 

\end{abstract}



\maketitle

\section{INTRODUCTION}

The consistency of the Higgs boson ~\cite{atlasH, cmsH} with the standard model has been tested extensively in many different channels. 
Importantly, in 2018, the coupling of the Higgs boson with the top quark was observed in top quark pair production in association with the Higgs boson ($\ttH$)~\cite{ttHCMS,ttHATLAS} directly.
As the branching fraction of the Higgs boson to $\bbbar$ is the largest,
the $\ttH(b\bar{b})$ process can be measured with the best statistical precision. 
However, understanding the $\ttbb$ process precisely is essential for improving 
the sensitivity of a search for the $\ttH(b\bar{b})$ process.

The theoretical next-to-leading-order (NLO) calculation of the $\ttbb$ process was done~\cite{ttbbQCD} in the same phase space where the inclusive cross-sections were measured at $\sqrt{s}$ = 8 TeV in the CMS experiment~\cite{ttbbCMS8TeV}. However, both theoretical and experimental results have large uncertainties.   
The inclusive cross-sections of the $\ttbb$ process have been measured in the dilepton channel, also at $\sqrt{s}$ = 13 TeV 
in the CMS experiment~\cite{ttbbCMS13TeV}. This analysis was updated with more data recently, including the lepton+jets channel~\cite{ttbbCMS13TeVupdate}. The cross-section in the hadronic channel was also measured~\cite{ttbbHad}. These measurements in different channels show consistently that the measured inclusive cross-sections are higher than the theoretical predictions. 
The inclusive and thedifferential $\ttbb$ cross-sections were measured in the ATLAS experiment~\cite{ttbbATLAS}.   
However, in the ATLAS measurement, the origin of the b jet was not identified.
We can provide more information to the theorists by identifying the additional b jet and measuring their 
differential cross-sections. 
In real data, this is very challenging because no single variable can distinguish between additional b jets and the b jets from top quark decays.
In the CMS experiment, using early data at $\sqrt{s}$ = 8 TeV, 
identifying the additional b jets was already attempted for the first time with a boosted decision tree (BDT) in the dilepton channel~\cite{ttbbdiffCMS8TeV}.
In Ref.~\cite{ttbbdiffCMS8TeV}, the measured differential cross-sections as a function of the $\pt$ and the $|\eta|$ of the leading and sub-leading additional b jets, $\Delta R(b,\bar{b})$ and the invariant mass m$(b,\bar{b})$ of two additional b jets were reported. That analysis showed that the kinematic reconstruction of the $\ttbar$ process did not help improve correct assignments for the additional b jets. 
Recently, a deep neural network (DNN) has been proposed to reconstruct the $\ttbar$ events and compared with kinematic fitting~\cite{dnnfortopReco}. However, that paper did not attempt to match the additional b jets in the $\ttbb$ events. 

In this paper, the identification of two additional b jets in the $\ttbb$ process in the lepton+jets channel by using the minimum $\Delta R(b,\bar{b})$ method, BDT and DNN techniques is reported. The performances of these approaches are compared quantitatively.

\begin{figure}[h]
\includegraphics[width=6cm]{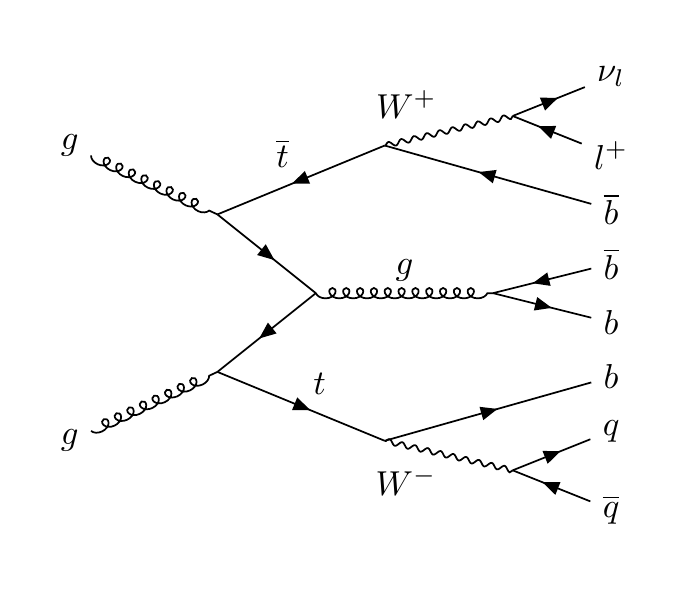}
\caption{
Feynman diagram for the $\ttbb$ process in the lepton+jets where one of the W bosons decays hadronically.
}\label{fig:feynttbb}
\end{figure}

The Feynman diagram for the $\ttbb$ process in the lepton+jets mode is shown in Fig.~\ref{fig:feynttbb}.
This measurement identifying additional b jets from gluon splitting suffers from large combinatorial backgrounds. For example, if 6 jets are in the clean $\ttbb$ events, 
the probability of identifying additional b jets with random choice is only around 7\% when two additional jets are known to be in the selected events. 
For this reason, the dilepton channel would have an advantage compared to the lepton+jets channel.
However, if the differential cross-section of $\ttbb$ is to be measured,
having larger statistics provided by the lepton+jet channel is crucial. 
Therefore, this study makes use of the lepton+jet channel, which has a larger cross-section.

This study is focused on finding two additional b jets from the gluon splitting in the lepton+jet channel by using different jet assigning methods for quantitative comparisons. This study will provide valuable information towards precise differential cross-section measurements in the lepton+jet channel.  

\section{Simulation}\label{sec:sam}
The simulated $\ttbb$ events in pp collisions are produced at a center-of-mass energy of 13 TeV.
We generated 18M events for the $\ttbb$ samples by
using the {\sc MadGraph5}\_aMC@NLO program (v2.6.6)~\cite{aMCNLO} at the leading order, and these events are further interfaced to  {\sc Pythia} (v8.240)~\cite{pythia} for the hadronization. 
A W boson decays through {\sc MadSpin}~\cite{madspin}, and the events are generated in a 4-flavor scheme, where the b quark has mass.

The generated events are processed by using the detector simulation with the DELPHES package (v3.4.1)~\cite{delphes} for the CMS detector. 
The physics objects used in this analysis are reconstructed based on the particle-flow algorithm~\cite{pfa} implemented in the DELPHES framework.
In the DELPHES fast simulation, the final momenta of all the physics objects,
such as electrons, muons and jets, are smeared as a function of the transverse momentum $\pt$ and the pseudorapidity $\eta$ so that
they can represent the detector effects. 
The efficiencies of the identification of the electrons, muons and jets are also parameterized as functions of $\pt$ and $\eta$
based on information from the measurements made by using the CMS data~\cite{delphes}. 
The muon identification efficiency is set to 95\% for muons with 
momenta $\pt$ $>$ 10 GeV and $\pt$ $<$ 100 GeV.
The electron identification efficiency is set to 95\% for $|\eta|$ $>$ 1.5 and 85\% for 1.5 $<$ $|\eta|$ $<$ 2.5.   
The isolated muons and electrons are selected by applying a relative isolation of $I_{rel}$ $<$ 0.25 and 0.12, respectively, where $I_{rel}$ 
is defined as the sum of the surrounding energy from the particle-flow tracks, photons and neutral hadrons 
divided by the transverse momentum of the muon or the electron.  
The particle-flow jets used in this analysis are reconstructed by using the anti-{\it k}$_{\rm T}$ algorithm~\cite{antikt} to cluster the particle-flow tracks and particle-flow towers. 
If the jet is already reconstructed as an isolated electron, muon or photon, it is excluded from further consideration.  
The b-tagging efficiency is precisely parameterized as a function of $\pt$ and $\eta$ of the jet by using polynomial functions:
\begin{equation}
\epsilon_b(\pt) = a_b+b_b\ \pt+c_b\ \pt^2+d_b\ \pt^3+e_b\ \pt^4 + f_b\ \pt^5,
\label{eq:btag}\end{equation}
where all the coefficients in the above polynomial are given in Table~\ref{tab:btag}. 

\begin{table*}[ht]
  \centering
  \small
  \begin{tabular}{ | c | c c c c c c |}
  \hline
  Jet $\pt$ (GeV) & $a_b$ & $b_b$ [GeV$^{-1}$]& $c_b$ [GeV$^{-2}$]
    & $d_b$ [GeV$^{-3}$] & $e_b$ [GeV$^{-4}$] & $f_b$ [GeV$^{-5}$]\\
  \hline
  \hline
  $ 20 - 50$ & $-0.033$ & $\phantom{-}0.0225$ & $-3.5\cdot10^{-4}$ & $\phantom{-}2.586\cdot 10^{-6}$ & $-9.096\cdot10^{-9}$ & $\phantom{-}1.212\cdot 10^{-11}$ \\
  $ 50 - 160$ & $\phantom{-}0.169$ & $\phantom{-}0.013$ & $-1.9\cdot10^{-4}$ & $\phantom{-}1.373 \cdot 10^{-6}$ & $-4.923\cdot 10^{-9}$ & $\phantom{-}6.87\cdot 10^{-12}$ \\
  $ 160 - 1000$ & $\phantom{-}0.62$ & $-0.00083$ & $\phantom{-}4.3078\cdot 10^{-7}$ & 0 & 0 & 0\\
  \hline 
  \end{tabular}
  \centering
  \caption{\label{tab:btag} Coefficients related to the parameterization of the
   b-tagging efficiencies associated with our fast detector simulation for
   the different working points. See Eq.~\eqref{eq:btag}.}
\end{table*}

The b-tagging efficiency is around 50\% at the tight-working point of the deep combined secondary vertex (DeepCSV) algorithm~\cite{deepjet},
which shows the best performance in the CMS measurement~\cite{cmsbjet}. 
The corresponding fake b-tagging rates from the c-flavor and the light flavor jet are set to around 2.6\% and 0.1\%, respectively.  

Once events are produced, the $\ttbb$ process is defined based on the particle-level jets 
obtained by clustering all final-state particles at the generator level.
A jet is considered as an additional b jet if the jet is matched to the last b quark not from a top quark 
within $\Delta R(j,q)= \sqrt{ {\Delta \eta(j,q)}^2+{\Delta \phi(j,q)}^2 }$ $<$ 0.5,
where j denotes jets at the generator level and q denotes the last b quark.  
The additional b jets are required to be within the experimentally accessible kinematic region of $|\eta|$ $<$ 2.5 and  $\pt$ $>$ 20 GeV.
At least two additional b jets should exist to be $\ttbb$ events. With this condition, we have 3397814 $\ttbb$ events which is 18\% of the generated sample. 

\section{Event Selection}\label{sec:evt}
We applied the following event selection to remove the main backgrounds from the multi-jet events and W+jet events. 
At the reconstruction level of the lepton+jet channel, the event must have exclusively one lepton with $\pt$ $>$ 30 GeV and $\eta$ $<$ 2.4 at the preselection ({\bf S1}).
At this preselection, 1010684 events survived with an acceptance of 5.6\%. 
Jets are selected with a threshold of $\pt$ $>$ 30 GeV and $|\eta|$ $<$ 2.5.   
The $\ttbb$ event has the final state of 4 b jets and two jets from one of two W bosons in top quark decays. 
However, the detector acceptance and the efficiencies of the b jet tagging algorithms are not 100\%. 
Some of the $\ttbb$ events have fewer jets at the reconstruction level. 
Therefore, starting from a requirement of at least two jets, we split the events into 12 categories based on the number of the jets ({\bf S2}) and b-tagged jets ({\bf S3}) exclusively.  
The number of selected events and the corresponding acceptance for each event selection step are shown in Table~\ref{tab:event}. 
These numbers of events are used for the $\Delta R$, and for the DNN methods in the following sections. 

\begin{table}[h]
\begin{center}
 \small
 \centering
 \begin{tabular}{| c | c | c | c |}
 \hline
 $~\geq N_j$          ~& $~N_{b} ~$ &  ~S2, S3 ( $\geq N_{j}$, $N_{b}$ ) ~&~ Acceptance (\%)~ \\ 
 \hline\hline
 2 & \multirow{5}{*}{$=$~2}  & 330704 & 1.84 \\
 3 &                    &    328054 & 1.82 \\
 4 &                    &    305890 & 1.70 \\
 5 &                    &   241585 & 1.34 \\
 6 &                    &   147066 & 0.82 \\ \hline
 3 & \multirow{4}{*}{$=$~3} & 138011 & 0.77 \\
 4 &                    &   134912 & 0.75 \\
 5 &                    &   116884 & 0.65 \\
 6 &                    &   79164 & 0.44 \\ \hline
 4 & \multirow{3}{*}{$\geq$~4} & 27653 & 0.15 \\
 5 &                    &  26413 & 0.15 \\
 6 &                    &  20653 & 0.11 \\
 \hline
 \end{tabular}
 \caption{
Number of events in each event selection based on the numbers of jets ($N_{j}$) and b-tagged jets ($N_{b}$) exclusively and the corresponding acceptances. With these numbers of events, the statistical uncertainty in the acceptance can be negligible. The additional b jets in the $\ttbb$ events are required to be within the experimentally accessible kinematic region of $|\eta|$ $<$ 2.5 and $\pt$ $>$ 20 GeV.
}
 \label{tab:event}
\end{center}
\end{table}

Two jets will be selected as candidates for additional b jets from either the $\Delta R$ or the DNN methods. 
If the selected jet matches any of the additional b jets at the generation level within $\Delta R < 0.4$, 
the jet is considered to be an additional b jet. 
Both selected jets should be matched to be a ``$\it{matched}$'' event. 
Then, the matching efficiency is defined as fraction of matched events in the category. 
The matching efficiency is used as a figure of merit to check the performance in this analysis.  

\begin{figure}
\includegraphics[width=6cm]{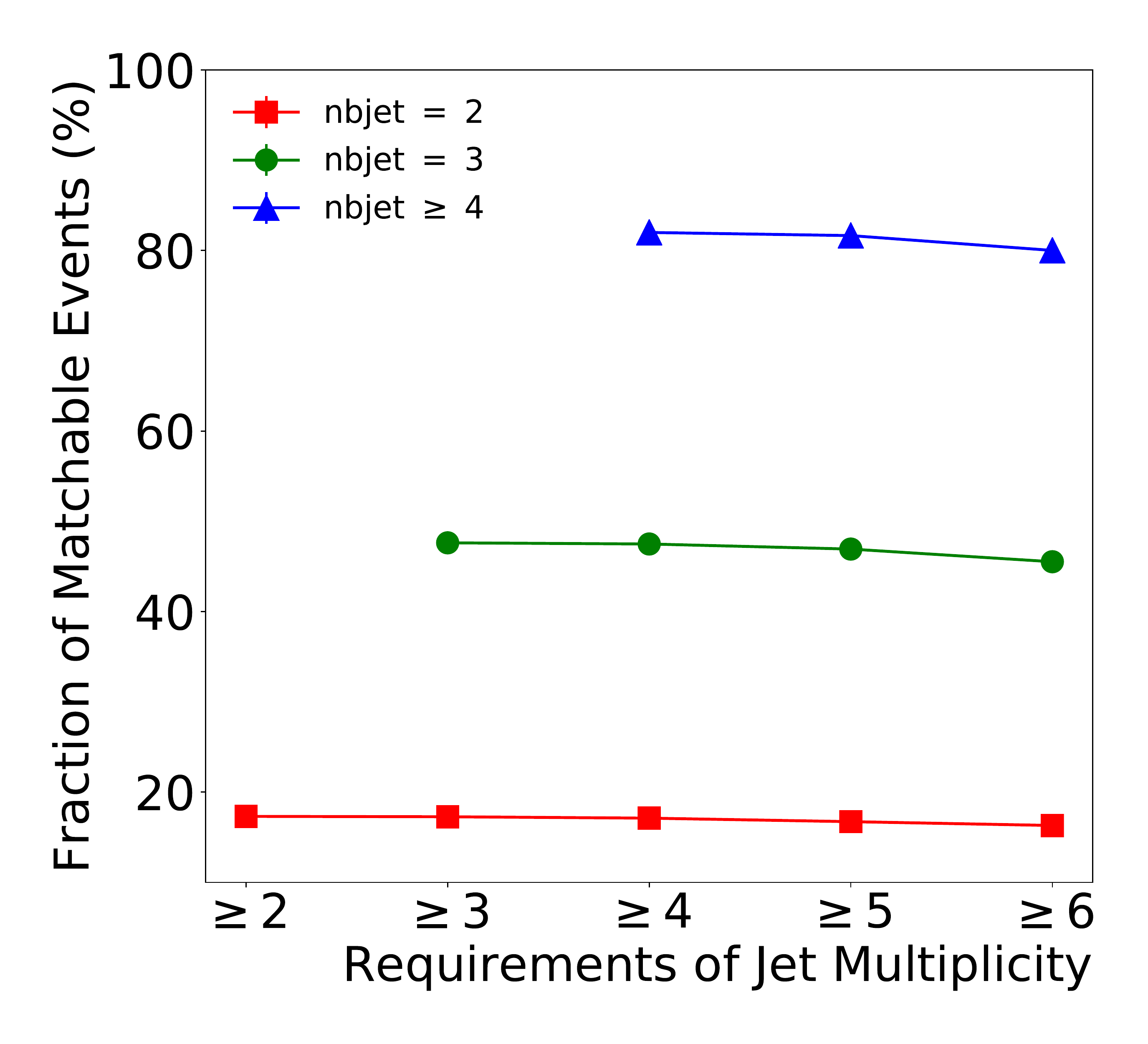}
\caption{
(color online) Fractions of matchable events in the selected dataset as functions of the number of jets.
The results for 2, 3, and 4 b-tagged jets are shown using 
rectangles, circles, and triangles, respectively.  
}\label{fig5:matchable}
\end{figure}

Figure~\ref{fig5:matchable} shows the fraction of selected events
with two additional jets that can be matched in
each selection. This plot has two aspects. 
As the jet multiplicity is higher, we tend to have more fake b-tagged jets, so the fraction of matchable events decreases. 
On the contrary, if we have more b-tagged jets, the chances of having the correct assignment are mmore. 
As expected, the ratio significantly goes up with the requirement of more b-tagged jets.  

\section{Minimum $\Delta R$ Analysis}\label{sec:minR}

\begin{figure}
\includegraphics[width=6cm]{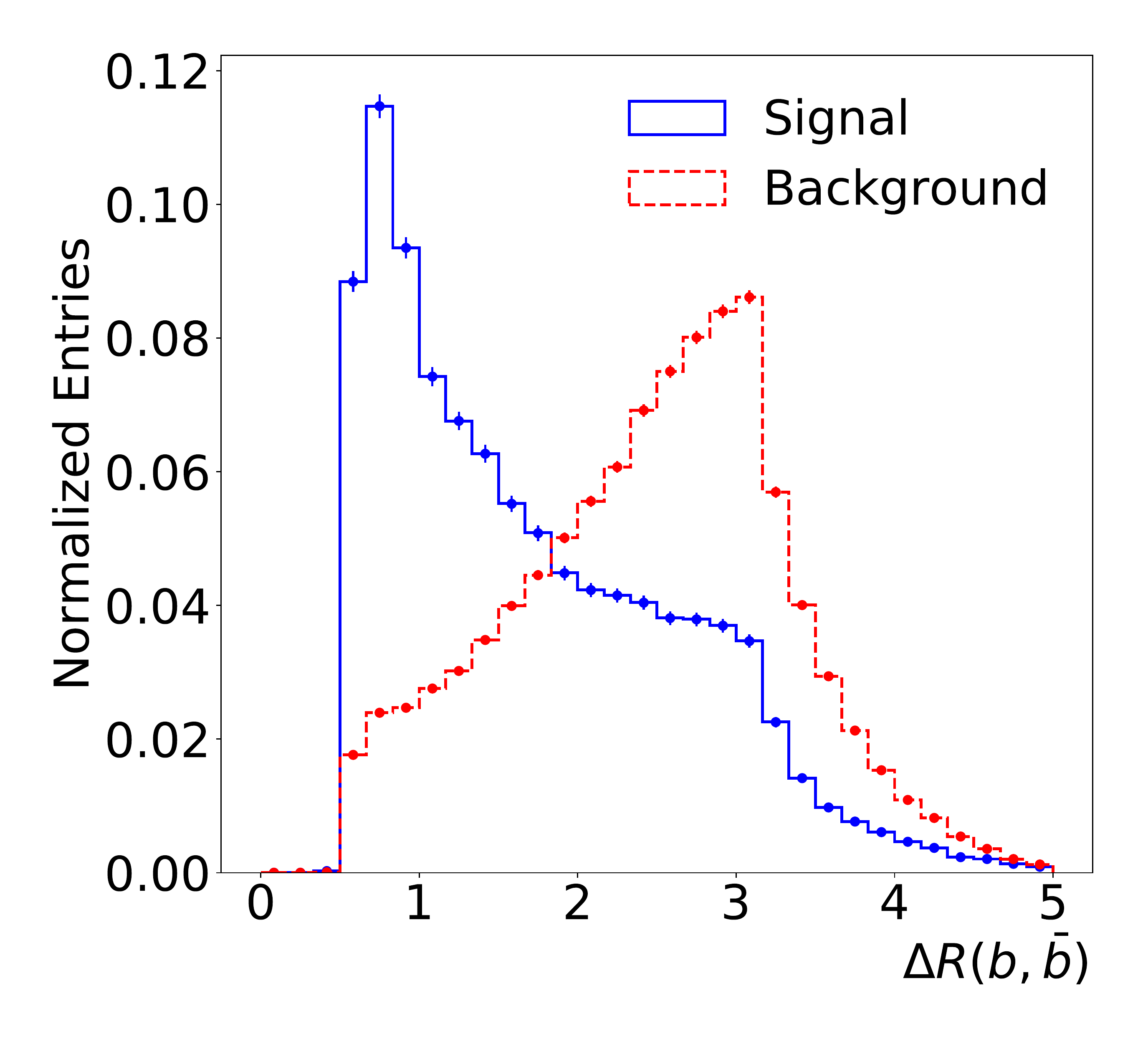}
\caption{
$\Delta R$ of two selected b jets for the correct assignment (signal) and the wrong assignment (background).
}\label{fig3:mindR}
\end{figure}

One simple and straightforward approach for identifying two additional b jets is to use a minimum angle of $\Delta R$ between them. 
This method is based on the fact that the additional b jets from the gluon splitting tend to have a smaller angle. 
The correct combination tends to have a smaller angle between two jets, so it can be distinguished from the wrong combinations (see Fig.~\ref{fig3:mindR}). 

\begin{figure}
\includegraphics[width=6cm]{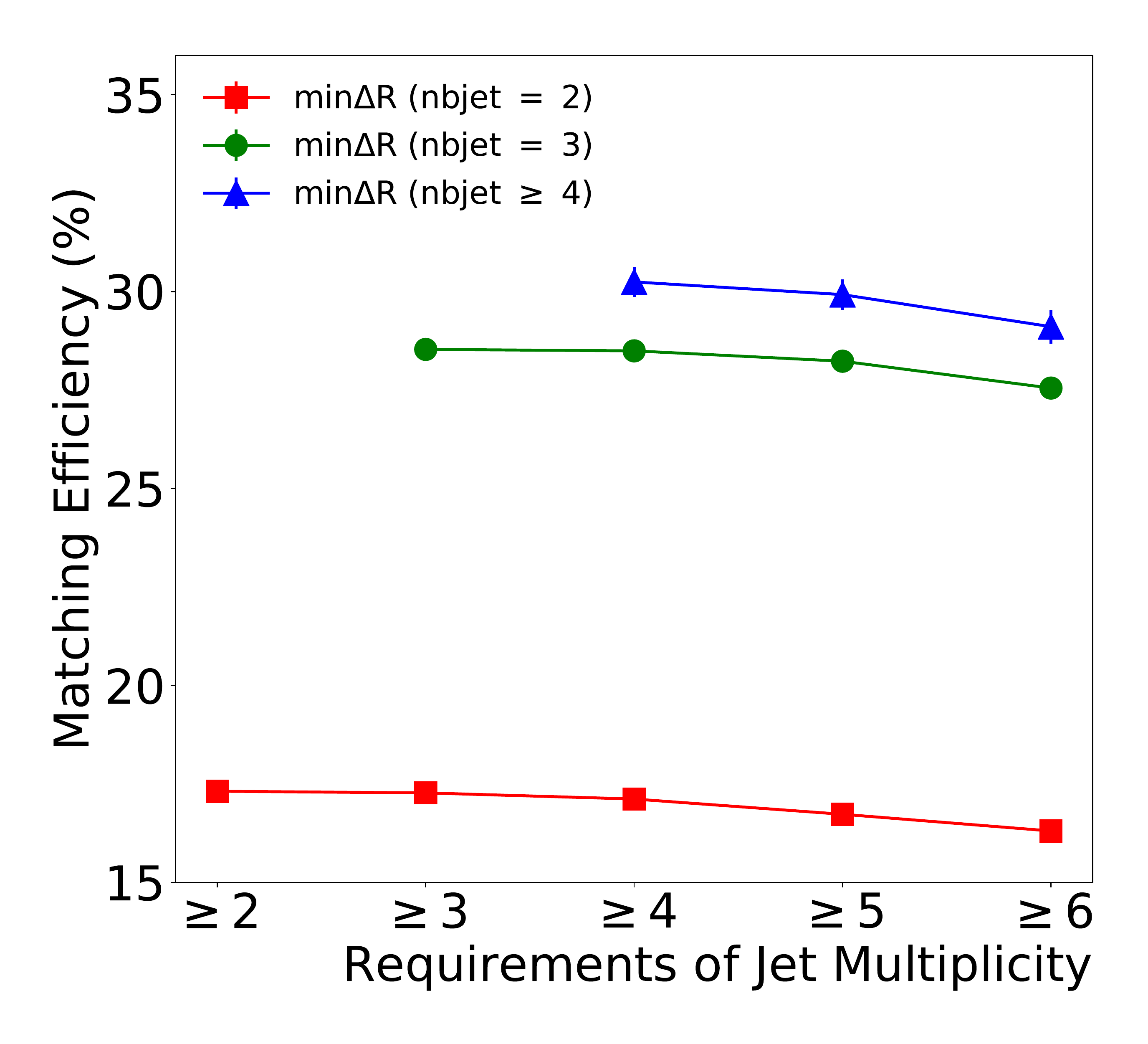}
\caption{
(color online) Matching efficiencies from the $\Delta R$ analysis are shown as functions of the number of jets. The results for 2, 3 and 4 b-tagged jets are shown, respectively, as rectangles, circles, and triangles. 
}\label{fig:minRreco}
\end{figure}

After the preselection, events are selected by requiring at least two jets and at least two b-tagged jets.
With requirements on various (b) jet multiplicities, the performance is tested in each category 
as the matching efficiency can be different. 
Figure~\ref{fig:minRreco} shows the matching efficiency as a function of the number of jets and b-tagged jets with different colors. 
The matching efficiency is calculated as around 17\% with the requirement of exclusively two b-tagged jets.
With the $\Delta R$ approach, the matching efficiency is 28\% for the requirement of 3 b-tagged jets and 30\% for at least 4 b-tagged jets.
In all of these cases, the matching efficiencies becomes slightly worse as more jets are required.
The results are compared together with the one using DNN in the following section. 

\section{Multivariate  Analysis}\label{sec:dnn}
As the second approach, the DNN method with multi-variables is used to increase the matching efficiency.
The goal of the neural network is to make use of multi-variables from the properties of the selected objects to decide which combination is most probable to have originated from gluon splitting.
The variables are selected considering all possible combinations of the four-vectors of the final state objects such as
selected two b-tagged jets, a lepton, a reconstructed hadronic W boson and missing transverse energy (MET), which are considered low-level features. 
Lists of the total 78 variables are shown in Tables~\ref{tab:input1} and~\ref{tab:input2}.

\begin{table}[t]
\fontsize{10}{7.2}\selectfont
\begin{center}
 \small
 \centering
 \begin{tabular}{| c | c | c |}
 \hline
 Variables & Description & Objects $(x,y)$\\
 \hline\hline
 $\Delta$R$(x,y)$ & $\Delta$R between $x$ and $y$ & \multirow{17}{*}{\shortstack{(b$_i$,b$_j$), \\ (b$_i$b$_j$,${\rm l}$), (b$_i$b$_j$,$\nu$), \\ (b$_i$,${\rm l}$), (b$_i$,$\nu$),\\ \\(b$_j$,${\rm l}$), (b$_j$,$\nu$), \\(b$_i$,W$_{\rm h}$), (b$_j$,W$_{\rm h}$)}}\\
 $\Delta\eta(x,y)$ & $\Delta\eta$ between $x$ and $y$ & \\
 $\Delta\phi(x,y)$ & $\Delta\phi$ between $x$ and $y$ & \\
 $\pt(x,y)$ & $\pt$ of $x$ and $y$ & \\
 $\eta(x,y)$ & $\eta$ of $x$ and $y$ & \\
 $m(x,y)$ & Invariant mass of $x$ and $y$ & \\
 $\mt(x,y)$ & Transverse mass of $x$ and $y$ & \\
 $H_T(x,y)$ & Scalar sum of $\pt$ of $x$ and $y$ & \\
 \hline
 \end{tabular}
 \caption{Input variables of 72 $(8\times9)$ from the two objects for DNN. The symbols b$_{i}$ and b$_{j}$ indicate two b-tagged jets from each combination, with $l$ indicating the selected lepton and $\nu$ indicating the MET. ${\rm W}_{\rm h}$ indicates the hadronic decay W boson.}
 \label{tab:input1}
\end{center}
\end{table}

\begin{table}[t]
\fontsize{10}{7.2}\selectfont
\begin{center}
 \small
 \centering
 \begin{tabular}{| c | c |}
 \hline
 Variables & Objects (b$_i$, b$_j$)\\
 \hline\hline
 $\pt$ & \multirow{5}{*}{\shortstack{~each of two b-tagged jets~}}\\
 $\eta$ & \\
 energy & \\
 \hline
 \end{tabular}
 \caption{Input variables of 6 $(3\times2)$ from the two b-tagged jets for DNN. The symbols b$_{i}$ and b$_{j}$ indicate two b-tagged jets from each combination. The signal is defined as the correct combination and the background as the wrong combination.}
 \label{tab:input2}
\end{center}
\end{table}

\begin{figure}
\includegraphics[width=6cm]{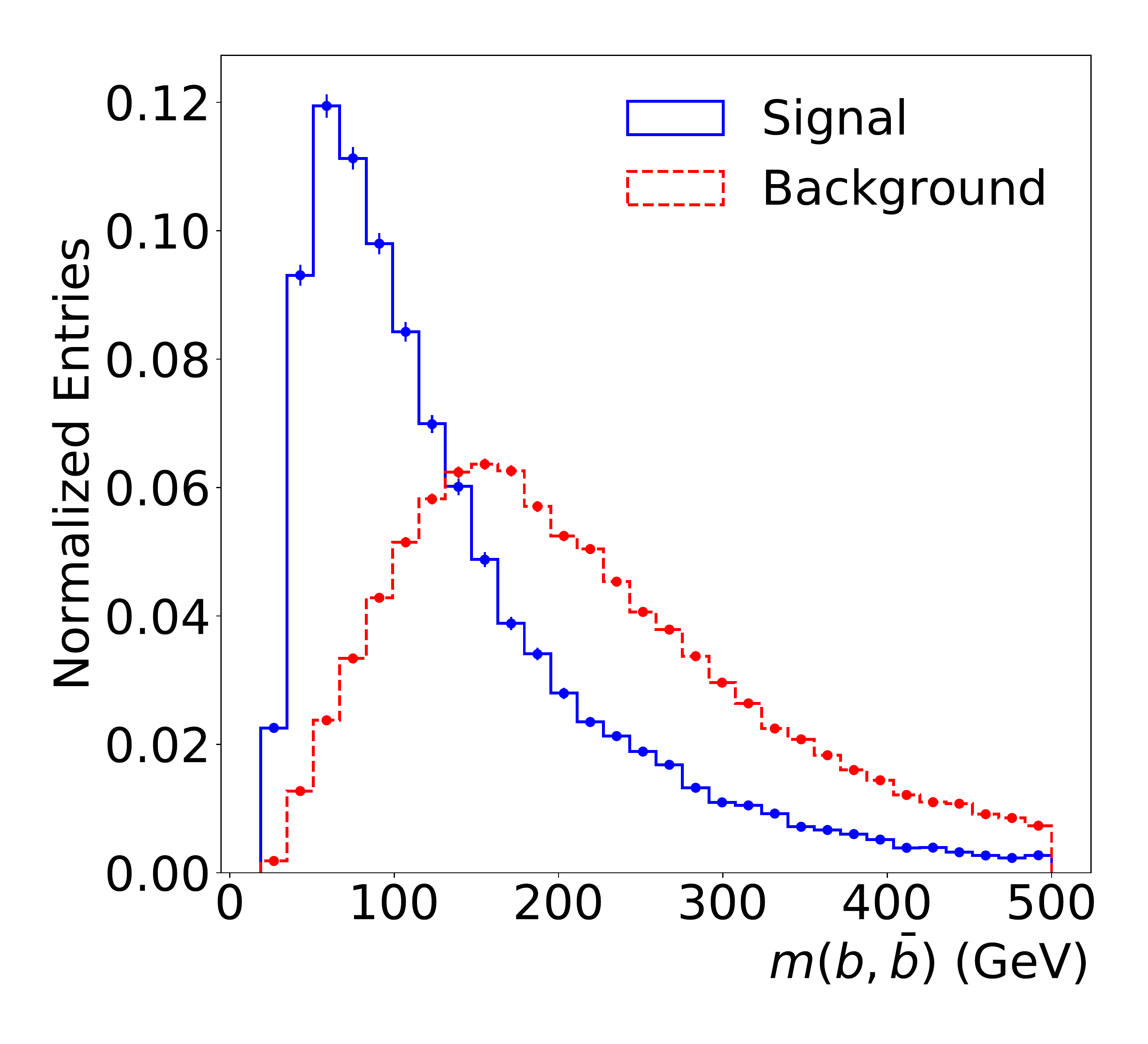}
\includegraphics[width=6cm]{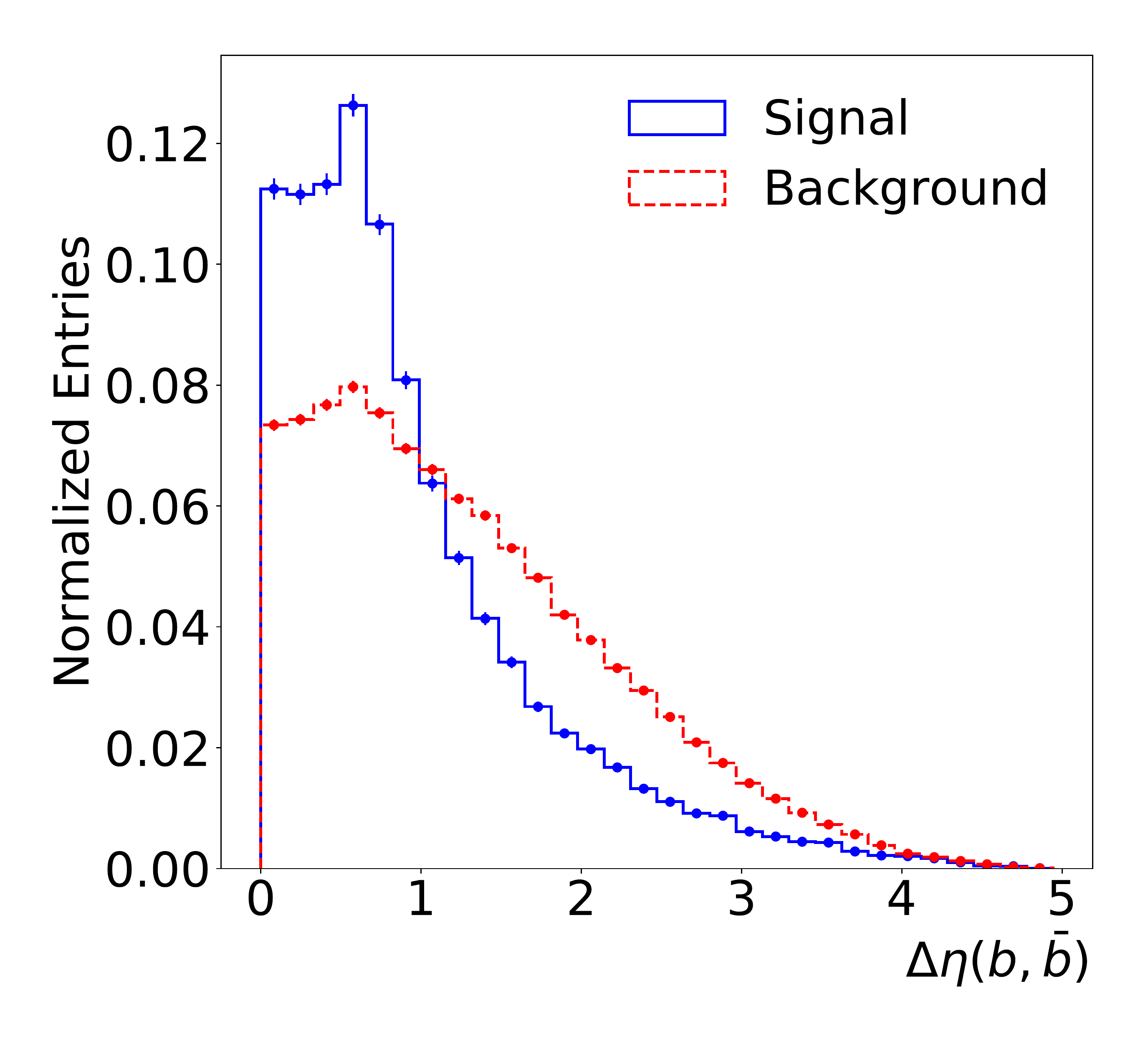}\\
\includegraphics[width=6cm]{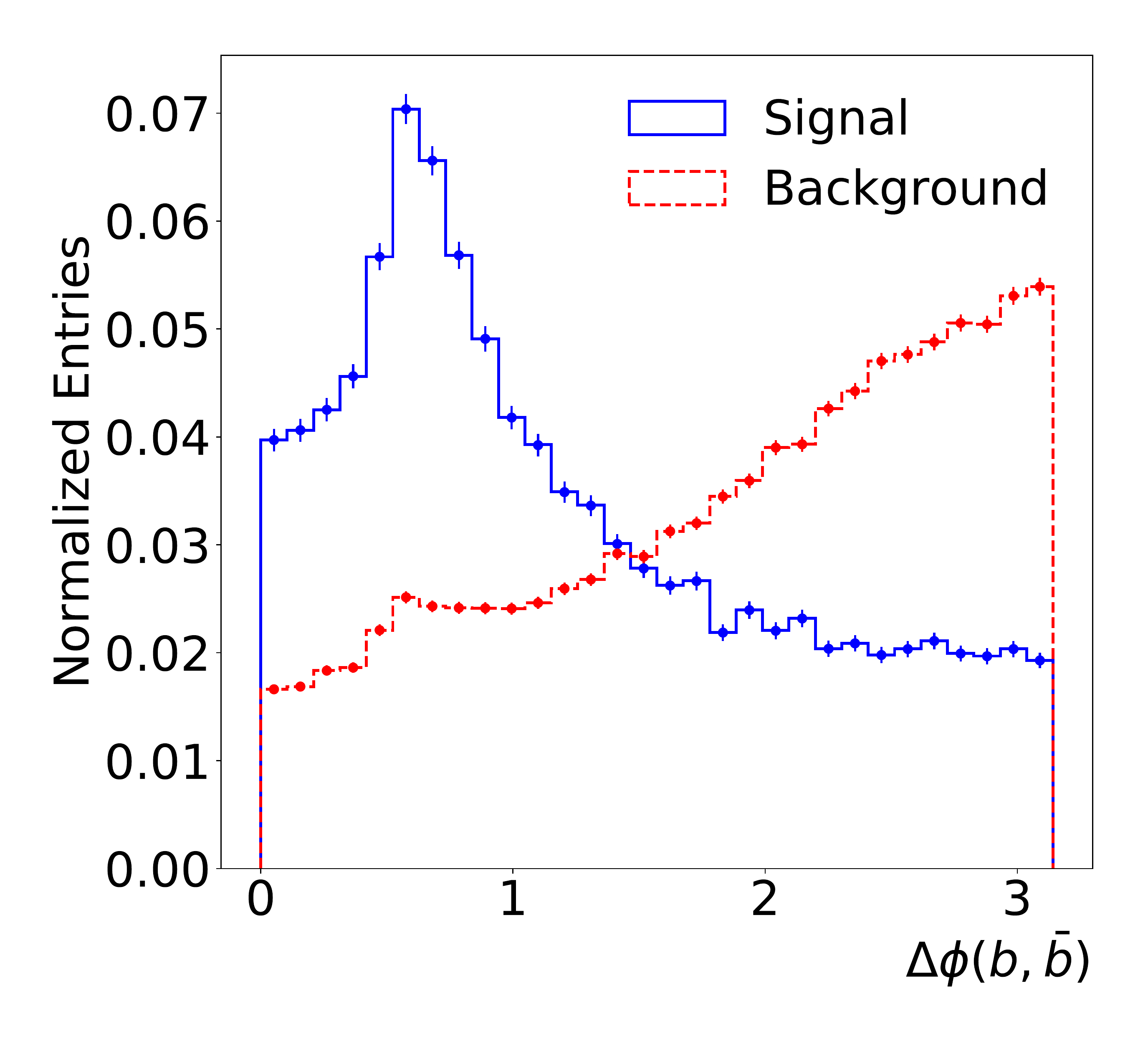}
\includegraphics[width=6cm]{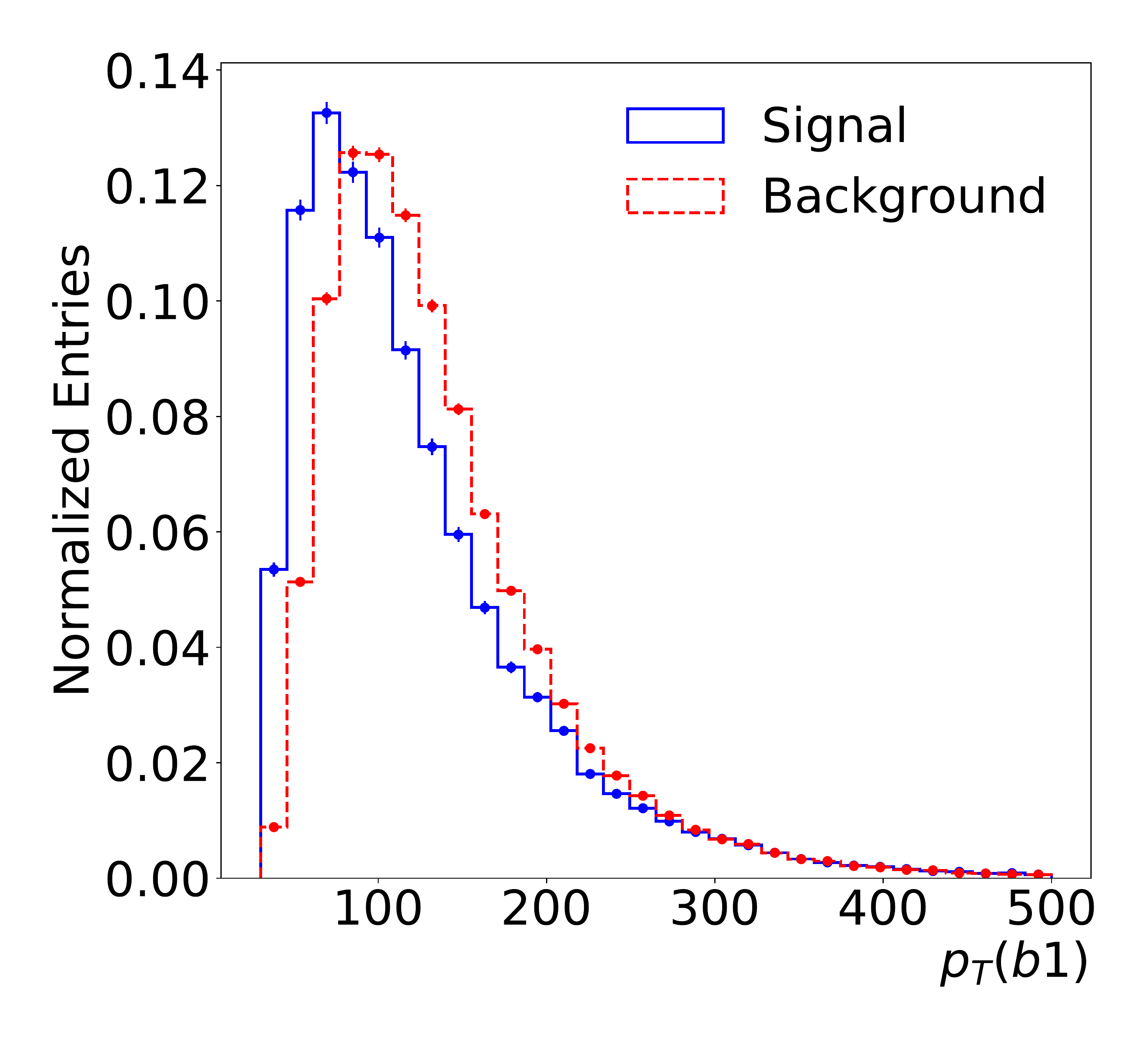}
\caption{
Invariant mass, $\Delta \eta$ and $\Delta \phi$ of the two selected b jets and $\pt$ of the selected leading b jet.
}\label{fig5:input}
\end{figure}

We will use the DNN 
implemented in Keras~\cite{keras} as feedforward neural networks
for the binary classification. For this classification, if two selected jets among the combinations of all b-tagged jets are matched to additional b jets, this combination is considered as ``$\it{signal}$''. 
All other combinations are considered as ``$\it{background}$''. 
For example, for 3 b-tagged jets, 3 combinations of two jets are possible. 
Possibly, one combination is a signal, and the other two combinations are backgrounds. 
Therefore, requiring more jets implies more background in training.
The signal and the background distributions of the most easily distinguished variables are shown in Fig.~\ref{fig5:input}. 

\begin{figure}[h]
\includegraphics[width=8cm]{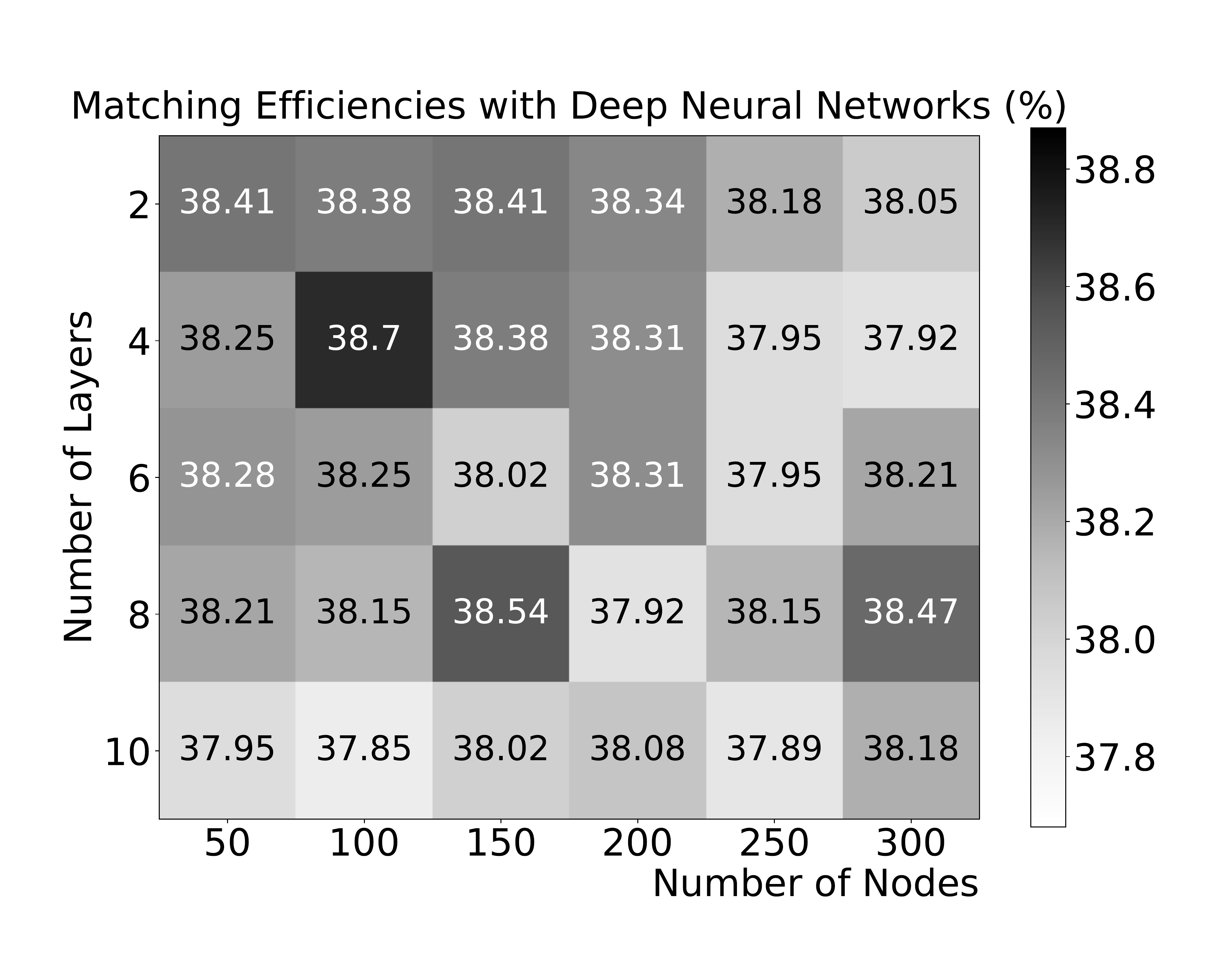}
\caption{
Matching efficiency in 2D for different numbers of layers and nodes per each hidden layer ranges under the requirement of at least 4 jets and at least 4 b-tagged jets.
Four layers and 100 nodes are selected as the optimal numbers. 
}\label{fig5:2d}
\end{figure}

The dataset in each event selection is split into two datasets, 80\% of data to train the model and the remaining 20\% for test. 
The neural network's hyperparameters, such as the number of epochs, the number of layers, and the number of nodes per each hidden layer 
are optimized based on the matching efficiency calculated on the test dataset.  
To minimize overtraining, we use regularization technique of L2 regularization, batch normalization~\cite{batchnorm} and dropout method~\cite{dropout} dropping out nodes by 8\% in each hidden layer. 
Two main hyperparameters of the number of layers and the number of nodes in each hidden layer are configured 
by scanning the 2D parameter space within the number of epochs of 100.
The parameters are chosen in a way that the efficiency is the highest. 
This procedure is done for each event selection scenario.  
Figure~\ref{fig5:2d} shows the matching efficiency in the 2D parameter space for the number of nodes and layers 
after at least 4 jets and 4 b-tagged jets. We found that the efficiency is not sensitive in this parameter space.
Therefore, the number of 4 hidden layers and 100 nodes are fixed for all event categories, which are considered as optimal choices overall.  

\begin{figure}
\includegraphics[width=6cm]{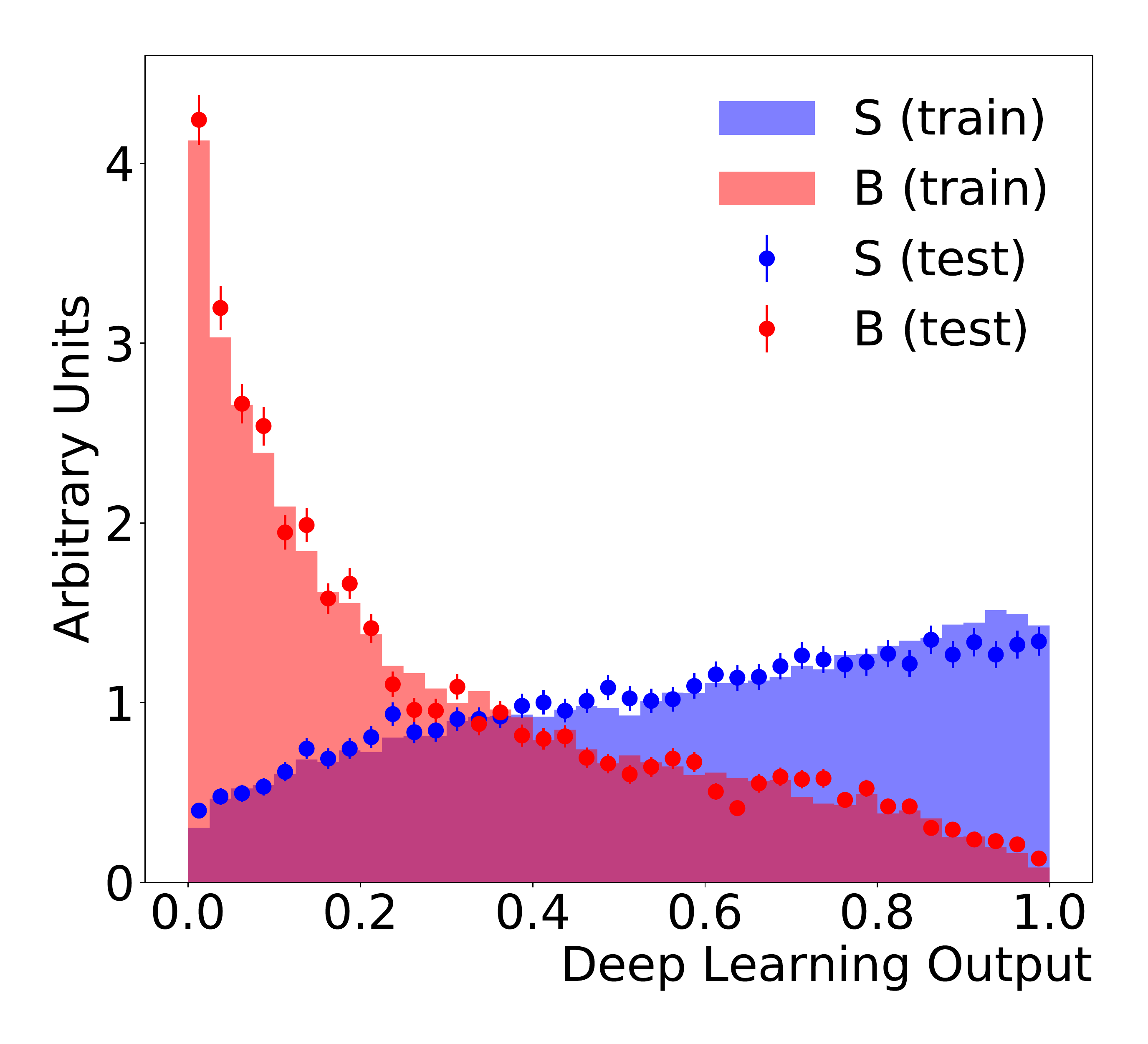}
\caption{
DNN output scores for training and test samples. 
The symbol S is the signal (correct combinations), and the symbol B is the background (wrong combinations).  
}\label{fig5:over}
\end{figure}

The output scores from the training and the test samples are shown in Fig.~\ref{fig5:over}. 
The distributions from both samples agree with each other. This agreement shows no overtraining in the model. For one event, average training time costs almost 0.1 seconds when using Nvidia Titan XP Graphic cards of 4 with 12 GB of RAM for the analysis.

\begin{figure}
\includegraphics[width=6cm]{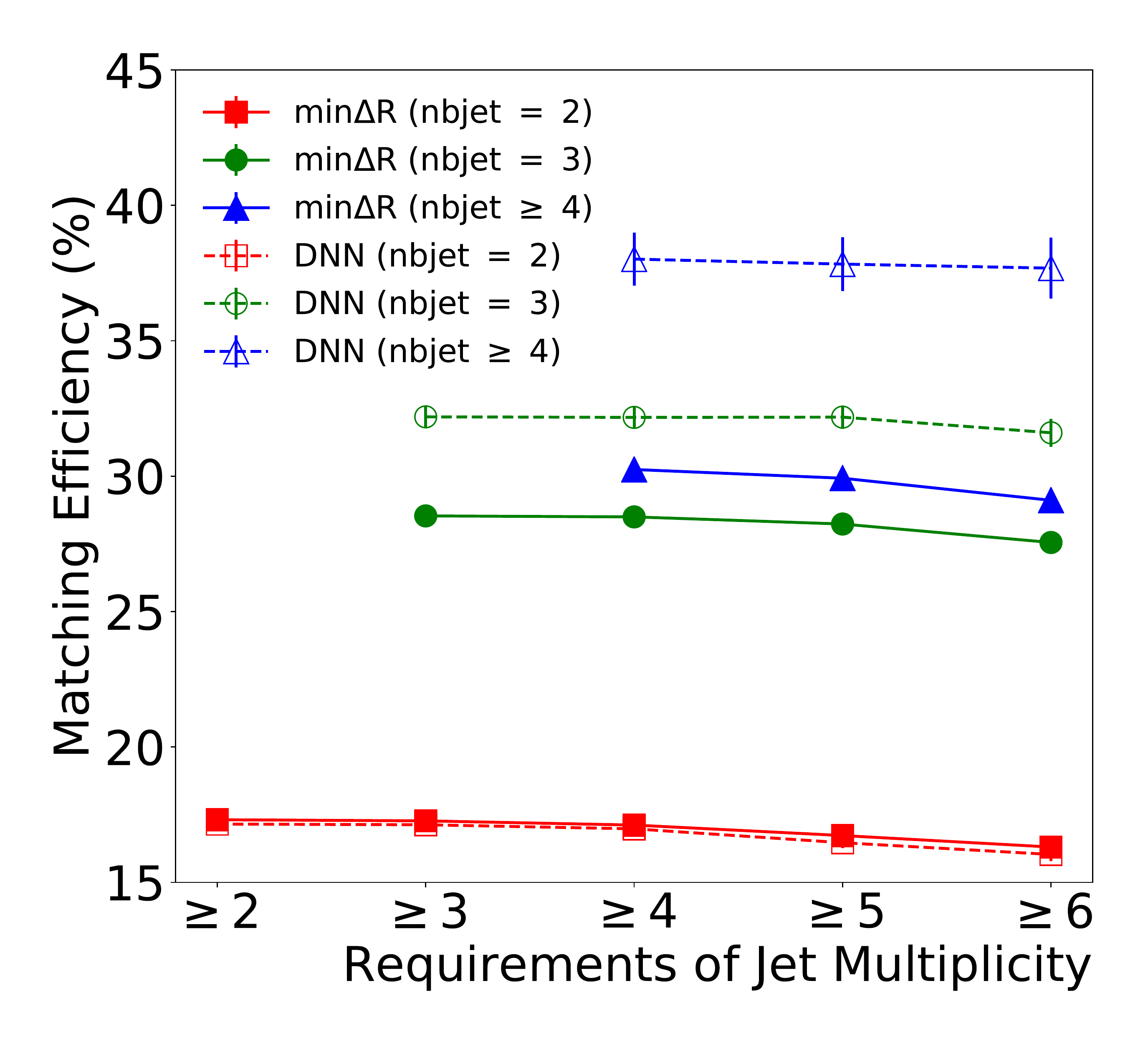}
\caption{
(color online ) The matching efficiency (dashed line) from the neural network comparison is shown together with the one from the minimum $\Delta R$ analysis (solid line) for various numbers of jets.
The results for 2, 3 and 4 b-tagged jets are shown, respectively, as rectangles circles, and triangles. 
}\label{fig5:matching}
\end{figure}

Figure~\ref{fig5:matching} shows the matching efficiencies for different requirements on the number of jets and b-tagged jets along with the results from the minimum $\Delta R$ analysis.
The corresponding numbers are shown in Table~\ref{tab:recoeff}.  
As shown in Fig.~\ref{fig5:matching}, overall, the matching efficiency is larger when a higher number of b-tagged jets is applied. 
After the requirement of exactly two b-tagged jets, no difference should exist between different jet assigning methods as we have only one pair of two b-tagged jets. 
For the requirement of 3 b jets, the matching efficiency of 32\% is obtained, which is around 3\% higher than that achieved using the $\Delta R$ method. 
With the requirement of at least 4 b jets, the matching efficiency goes up to 38\%. This efficiency is 8\% higher than that for the $\Delta R$ method.
In general, we could reach a matching efficiency level of around 40\%  
by using DNN in the lepton+jets mode. 

For completeness, a BDT within the TMVA framework~\cite{tmva} with the same input variables 
is also used in this study as another machine learning technique.
The BDT is tuned with the following parameters: number of trees of 100, and maximum depth of 3. 
The results are shown together in Table~\ref{tab:recoeff}.
The result from a DNN is comparative with the result from the BDT method. 

\begin{table}[ht]
\fontsize{10}{6.2}\selectfont
\begin{center}
 \small
 \centering
 \begin{tabular}{| c | c | c | c | c |}
 \hline
 \multirow{2}{*}{~$\geq N_j$~} & \multirow{2}{*}{~$N_{b}$~} & \multicolumn{3}{|c|}{Matching Efficiency (\%)}\\
 \cline{3-5}
 & &~ min.$\Delta$R ~&~ BDT ~&~ DNN ~\\ 
 \hline\hline
2	&	\multirow{5}{*}{$=$~2}	&	17.31	$\pm$	0.08	&	17.15	$\pm$	0.17	&	17.15	$\pm$	0.17	\\
3	&		&	17.27	$\pm$	0.08	&	17.13	$\pm$	0.17	&	17.13	$\pm$	0.17	\\
4	&		&	17.12	$\pm$	0.08	&	16.98	$\pm$	0.18	&	16.98	$\pm$	0.18	\\
5	&		&	16.73	$\pm$	0.09	&	16.47	$\pm$	0.20	&	16.47	$\pm$	0.20	\\
6	&		&	16.31	$\pm$	0.11	&	16.03	$\pm$	0.25	&	16.03	$\pm$	0.25	\\
\hline
3	&	\multirow{4}{*}{$=$~3}	&	28.53	$\pm$	0.16	&	32.11	$\pm$	0.39	&	32.19	$\pm$	0.39	\\
4	&		&	28.50	$\pm$	0.16	&	32.10	$\pm$	0.40	&	32.17	$\pm$	0.40	\\
5	&		&	28.23	$\pm$	0.18	&	32.14	$\pm$	0.43	&	32.18	$\pm$	0.43	\\
6	&		&	27.55	$\pm$	0.21	&	31.48	$\pm$	0.51	&	31.60	$\pm$	0.51	\\
\hline					
4	&	\multirow{3}{*}{$>=$~4}	&	30.25	$\pm$	0.38	&	37.56	$\pm$	0.97	&	38.02	$\pm$	0.97	\\
5	&		&	29.92	$\pm$	0.38	&	37.38	$\pm$	0.99	&	37.83	$\pm$	0.99	\\
6	&		&	29.11	$\pm$	0.43	&	37.05	$\pm$	1.11	&	37.68	$\pm$	1.12	\\
\hline
\end{tabular}
 \caption{
Matching efficiencies for min. $\Delta R$, BDT and DNN analyses for each event selection. Only statistical uncertainties are given.
}
 \label{tab:recoeff}
\end{center}
\end{table}

\section{$\ttH$ vs $\ttbb$}\label{sec:dis}
\begin{figure}[h]
\includegraphics[width=6cm]{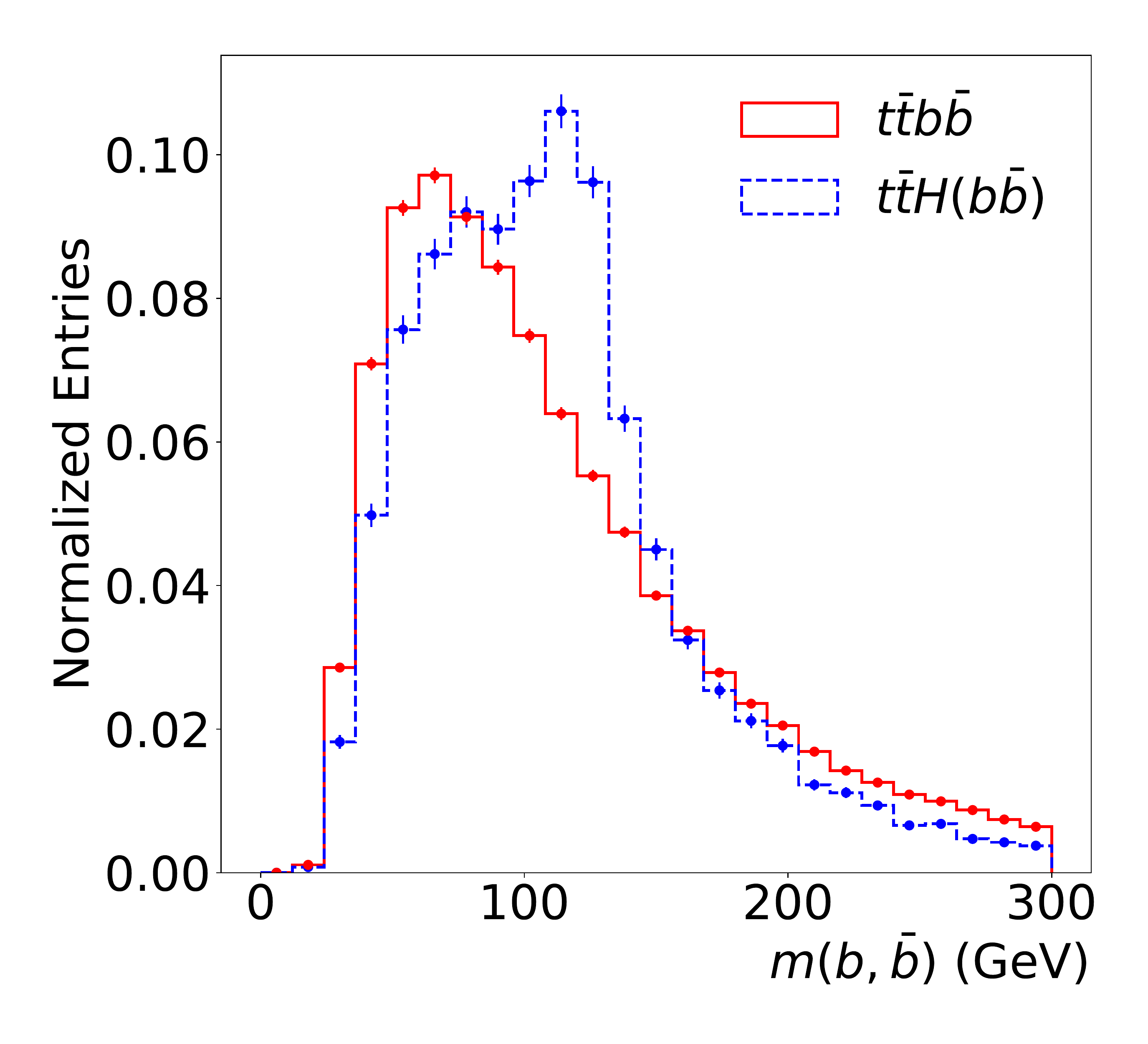}
\includegraphics[width=6cm]{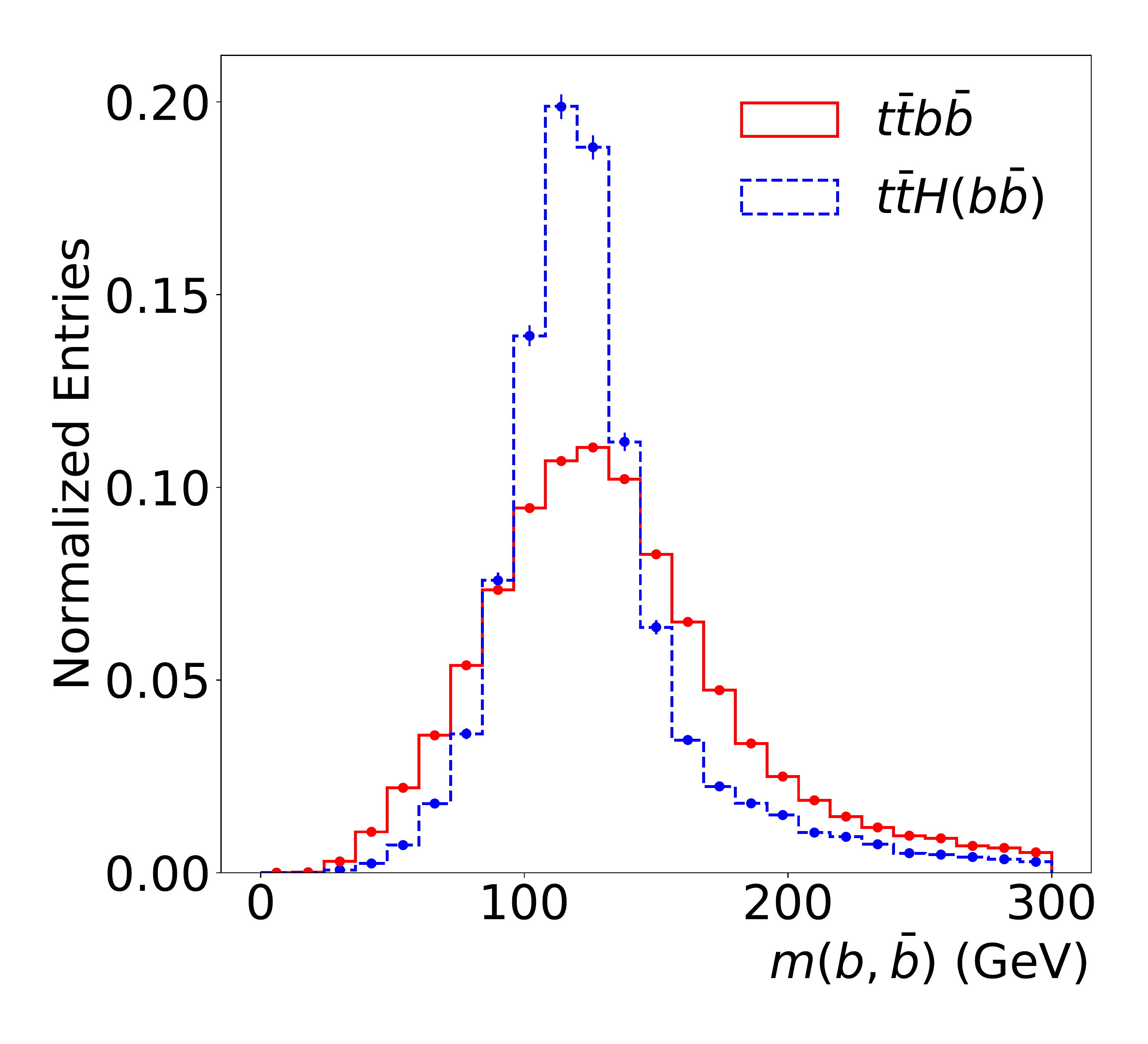}
\caption{
Comparisons at the reconstruction level between $\ttbb$ and $\ttH$ events for the distributions of the invariant mass from the model trained with $\ttbb$ events (left) and the model trained with $\ttH$ events (right).
}\label{fig:ttbbvstth}
\end{figure}

The DNN model trained with the $\ttbb$ events described above is used to identify two additional b jets, not from top quark decays. Because $\ttH$ events where the Higgs decays to $\rm b\bar{\rm b}$ also have additional b jets not from top quark decays, we can check the possibility to classify $\ttbb$ and $\ttH$ events by using the same model described above. Additionally, we trained the same model with the $\ttH$ events for possible improvement. Then, we compared the distributions of the invariant masses of two selected b jets from these models for $\ttbb$ and $\ttH$ events. Figure.~\ref{fig:ttbbvstth} shows the invariant mass spectra from the model trained with the $\ttbb$ events and the model with the $\ttH$ events after the requirement of at least 6 jets and exactly 3 b jets. The model trained with the $\ttH$ events yields a clear peak at the Higgs mass of 125 GeV for these events while it shows a broad peak for the model trained using the $\ttbb$ events. The two processes can still be distinguished. 

Usually, we separate these two steps: {\bf step 1.} identifying two additional jets and {\bf step 2.} classifying events with those selected jets. The second step involves another machine learning technique to improve performance. 
By extending the DNN model for multi-classification by using multiple processes, we should be able to tackle two things simultaneously: the identification of the origin of the b jets and the event categorization, such as for $\ttbb$ and $\ttH$ processes. 

\section{CONCLUSIONS}\label{sec:con}
If the differential cross-section are to be measured as functions of the properties of the additional b jets,
the origin of the b jet needs to be identified. 
In this paper, we present the performance of identifying two additional b jets in a top-quark pair production in the lepton+jets channel by using simulated pp collision data at $\sqrt{s}$ = 13 TeV.
Compared to the performance from the minimum $\Delta R$ analysis, the performance of DNN is improved by 3 - 8\% level, depending on the number of b-tagged jet required.
We also show that requiring at least 4 b-tagged jets leads to better performance, yielding a matching efficiency of around 40\%. 
This performance is about 8\% better than that obtained using the minimum $\Delta R(b,\bar{b})$ method. 
With BDT, the results are comparative overall and consistent within their statistical uncertainties. 

Requiring at least 4 b-tagged jets with a tight-working point of the b-tagging algorithm would be a promising event selection to increase the matching efficiency of identifying two additional b jets, especially when using DNN.    
However, this tight selection would suffer from the lack of statistics with the current LHC data.    
Using data from the Run-3 at the LHC or the High Luminosity-LHC in the future, a tightening of the requirement on the number of b-tagged jets to improve the matching efficiency would be conceivable. 
In a real analysis, more sophisticated high-level features such as b-tag discriminator are available and optimizing those variables can lead to improve the performance. 

\begin{acknowledgments}
This work was supported by the research fund of Hanyang University (HY-2015).
The work was also supported by the Basic Science Research Program through the National Research Foundation of Korea (NRF)
funded by the Ministry of Education, (Grant No. NRF-2020R1A2C2005228). 

\end{acknowledgments}

\end{document}